\documentclass[twocolumn,showpacs,preprintnumbers,superscriptaddress,amsmath,floatfix,amssymb,prl]{revtex4}
\usepackage[colorlinks=true]{hyperref}
\usepackage{graphicx}
\usepackage{epstopdf}
\newcommand{\comment}[1]{}

\newcommand{\lr}[1]{ \left( #1 \right) }
\newcommand{\lrs}[1]{ \left[ #1 \right] }

\newcommand{\vev}[1]{ \langle \, #1 \, \rangle }

\newcommand{\tr}{ {\rm Tr} \, }

\newcommand{\expa}[1]{ \exp{\left( #1 \right)} }

\newcommand{\logo}{\\ \vskip -18mm
\leftline{\includegraphics[scale=0.3,clip=false]{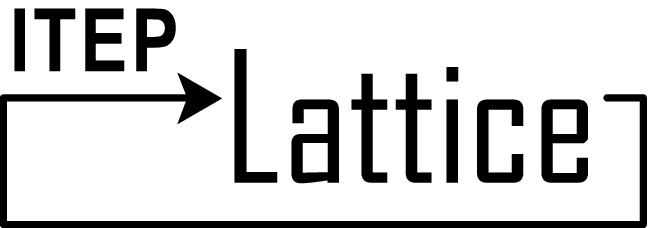}} \vskip 10mm}

\begin{document}
\sloppy
\preprint{ITEP-LAT/2010-12}

\title{Quark mass dependence of the vacuum electric conductivity induced by the magnetic field in $SU\lr{2}$ lattice gluodynamics\logo}

\author{P. V. Buividovich}
\email{buividovich@itep.ru}
\affiliation{ITEP, 117218 Russia, Moscow, B. Cheremushkinskaya str. 25}
\affiliation{JINR, 141980 Russia, Moscow Region, Dubna, Joliot-Curie str. 6}

\author{M. I. Polikarpov}
\email{polykarp@itep.ru}
\affiliation{ITEP, 117218 Russia, Moscow, B. Cheremushkinskaya str. 25}

\date{November 10, 2010}
\begin{abstract}
 We study the electric conductivity of the vacuum of quenched $SU\lr{2}$ lattice gauge theory induced by the magnetic field $B$ as a function of the bare quark mass $m$. The conductivity grows as the quark mass decreases. Simplest power-like fit indicates that the conductivity behaves as $B/\sqrt{m}$. We discuss the implications of this result for dilepton angular distributions in heavy ion collisions.
\end{abstract}
\maketitle

 Heavy ion experiments at RHIC have found an evidence \cite{Abelev:09:1} for the so-called Chiral Magnetic Effect (CME) \cite{Kharzeev:08:1} in quark-gluon plasma. The essence of the effect is the generation of electric current  along the direction of the external magnetic field in the background of topologically nontrivial gauge field configurations. Experimentally, the effect manifests itself as the dynamical enhancement of fluctuations of the numbers of charged hadrons emitted above and below the reaction plane in off-central heavy-ion collisions. In noncentral heavy-ion collisions, very strong magnetic field acting on hadronic matter is created due to the relative motion of the ions themselves.

 Recently, this enhancement of fluctuations was also shown to be related to the electric conductivity of the vacuum of the gauge theory \cite{Buividovich:10:1}. Namely, it was found that an external magnetic field applied to the confining vacuum of quenched non-Abelian lattice gauge theory induces nonzero conductivity in the direction of the field. In other directions, the vacuum remains an insulator. This phenomenon was called ``electric rupture by a magnetic field'', by analogy with similar phenomena known in condensed-matter physics \cite{Jonker:50:1}. Electric conductivity in the presence of an external magnetic field was also calculated in holographic models \cite{Lifschytz:09:1, Karch:07:1}.

 However, it turned out that theoretical and experimental relations between the CME and the conductivity induced by the magnetic field are far from trivial \cite{Shevchenko:10:1, Kharzeev:10:1}. In a recent paper \cite{Kharzeev:10:1}, it was suggested that massless excitation which is responsible for both the CME and the nonzero conductivity is the ``chiral spiral'', however, no qualitative predictions for the conductivity were obtained. In experiment, the induced conductivity is responsible for angular distributions of soft leptons rather than hadrons \cite{Buividovich:10:1} (see also \cite{Espriu:10:1}).

 Here we perform the simplest possible check of the ``chiral'' nature of the conductivity induced by the magnetic field: we study its dependence on the bare quark mass. As the nonzero quark mass leads to transitions between left- and right- handed states, if the imbalance of chirality is a necessary condition for the induced conductivity, it should decrease with the quark mass. Such decrease can, for example, serve to distinguish between ordinary and strange hadron matter produced in heavy-ion experiments \cite{Rafelski:82:1}.

 The technical setup of our simulations is the same as in \cite{Buividovich:10:1}. Namely, we extract electric conductivity from the Euclidean correlator of two vector currents $j_{i}\lr{x} = \bar{q}\lr{x} \gamma_{i} q\lr{x}$:
\begin{eqnarray}
\label{corr_def}
 G_{ij}\lr{\tau} = \int d^3 x \vev{ j_{i}\lr{0} j_{j}\lr{x, \tau}  }
 = \nonumber \\ =
\int \limits_{0}^{+\infty} \frac{d w}{2 \pi}\, K\lr{w, \tau} \rho_{ij}\lr{w} ,
\end{eqnarray}
where $\rho_{ij}\lr{w}$ is the corresponding spectral function, $K\lr{w, \tau} = \frac{w}{2 T}  \, \frac{\cosh{\lr{w \lr{\tau - \frac{1}{2 T}}}}}{\sinh{\lr{\frac{w}{2 T}}}}$ \cite{Aarts:07:1, Gupta:04:1}, $T = \lr{N_t a}^{-1}$ is the temperature, $a$ is the lattice spacing and $N_t$ is the size of the lattice in the time direction. The phenomenon of induced conductivity was found only in the confinement phase \cite{Buividovich:10:1}, therefore in the present paper we also limit our studies to the case of small temperatures.

 The underdefined linear system (\ref{corr_def}) is solved using the Maximal Entropy Method \cite{Asakawa:01:1, Aarts:07:1}. The Kubo formula for the electric conductivity yields the conductivity in terms of the spectral function in the limit of zero frequency \cite{Kadanoff:63:1, Aarts:07:1}:
\begin{eqnarray}
\label{Kubo_formula}
\sigma_{ij} = \lim_{\omega \to 0} \frac{\rho_{ij}\lr{\omega}}{4 T}\,.
\end{eqnarray}

 The correlator $\vev{ j_{i}\lr{x} j_{j}\lr{y}  }$ is measured on the equilibrium ensemble of quenched $SU\lr{2}$ gauge fields with lattice tadpole-improved action:
\begin{eqnarray}
\label{four_fermion_vev}
\vev{ j_{i}\lr{x} j_{j}\lr{y}  } =
\vev{\bar{q}\lr{x} \gamma_{i} q\lr{x} \, \bar{q}\lr{y} \gamma_{j} q\lr{y}}
= \nonumber \\ =
\int \mathcal{D}A_{\mu}\,e^{-S_{YM}\lrs{A_{\mu}}}
\nonumber \\
\tr\lr{\frac{1}{\mathcal{D}\lrs{A'_{\mu}} + m} \, \gamma_{i} \, \frac{1}{\mathcal{D}\lrs{A'_{\mu}} + m} \, \gamma_{j}},
\end{eqnarray}
where $A_{\mu}$ is the non-Abelian gauge field with the action $S_{YM}\lrs{A_{\mu}}$ and $\mathcal{D}\lrs{A_{\mu}}$ is the lattice Dirac operator. For $\mathcal{D}\lrs{A_{\mu}}$ we use Neuberger's chirally invariant overlap Dirac operator \cite{Neuberger:98:1}. We implement the Shifted Unitary Minimal Residue method (SHUMR) described in \cite{Borici:06:1} to find the quark propagator. For each value of the quark mass and the magnetic field we have used 30 configurations of gauge fields on $14^4$ lattice with lattice spacing $a = 0.102 \, fm$. In order to study the finite-volume effects, we have also repeated some calculations for $50$ configurations on the $20^3 \times 14$ lattice with the same lattice spacing. Uniform magnetic field is added to the Dirac operator by substituting $su\lr{2}$-valued vector potential $A_{\mu}$ with $u\lr{2}$-valued one $A'_{\mu \, ij} = A_{\mu \, ij} + 1/2\: F_{\mu\nu}\: x_{\nu} \delta_{ij}$. The magnetic field is directed along the $z$ axis. In order to account for periodic boundary conditions we introduce an additional boundary twist for fermions \cite{Wiese:08:1, Buividovich:09:7}.

\begin{figure*}[h]
  \includegraphics[width=5cm, angle=-90]{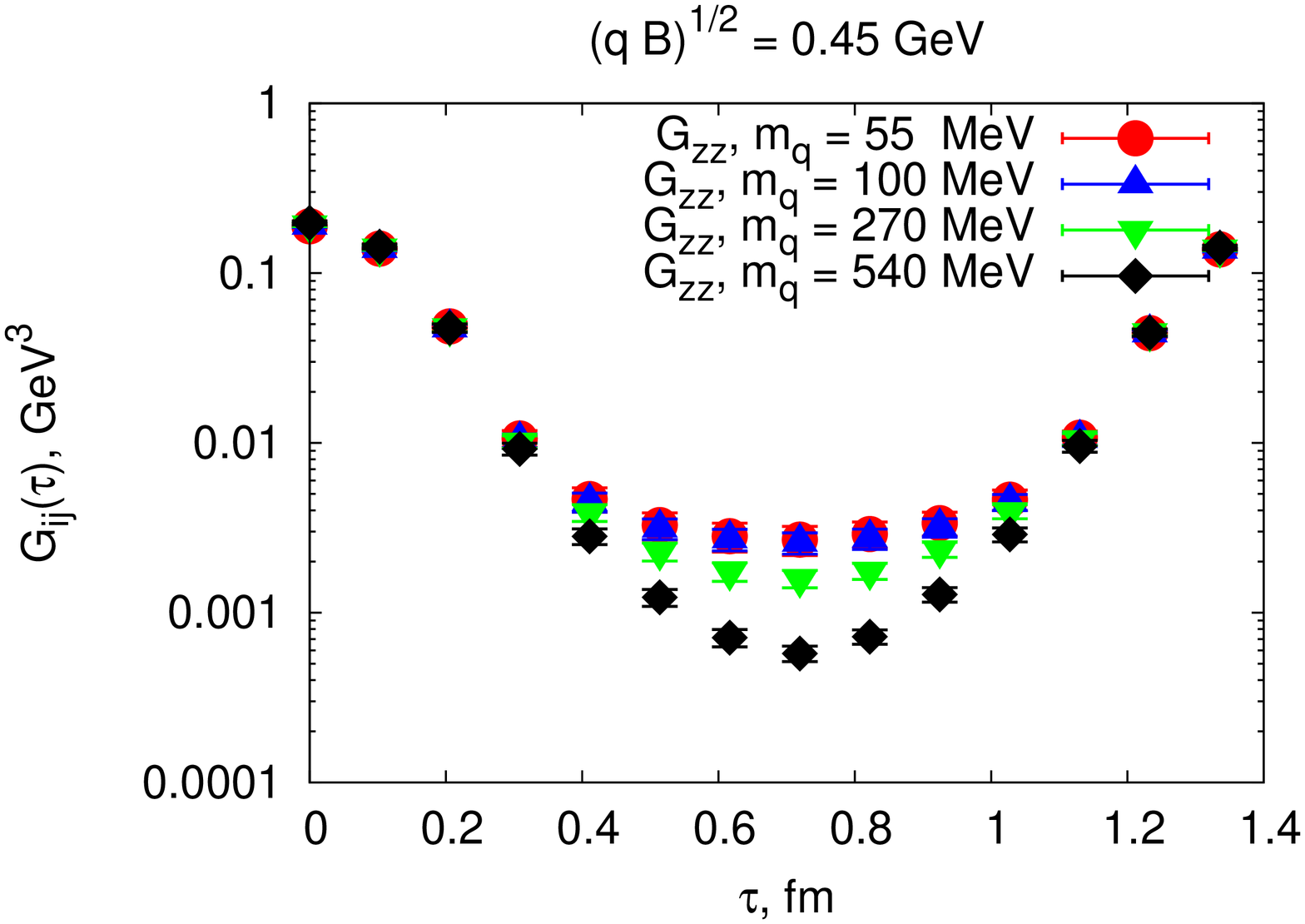}
  \includegraphics[width=5cm, angle=-90]{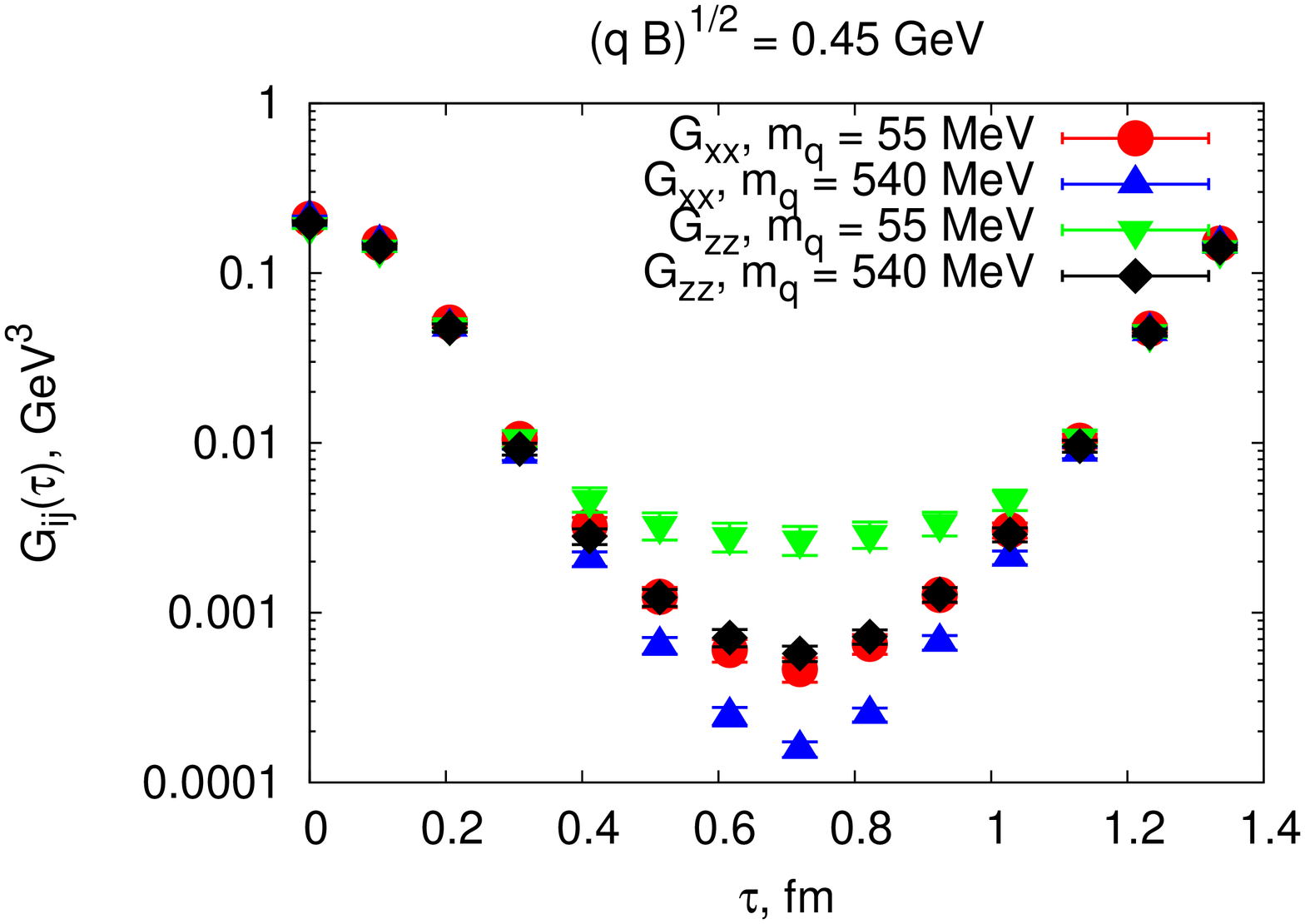}\\
  \includegraphics[width=5cm, angle=-90]{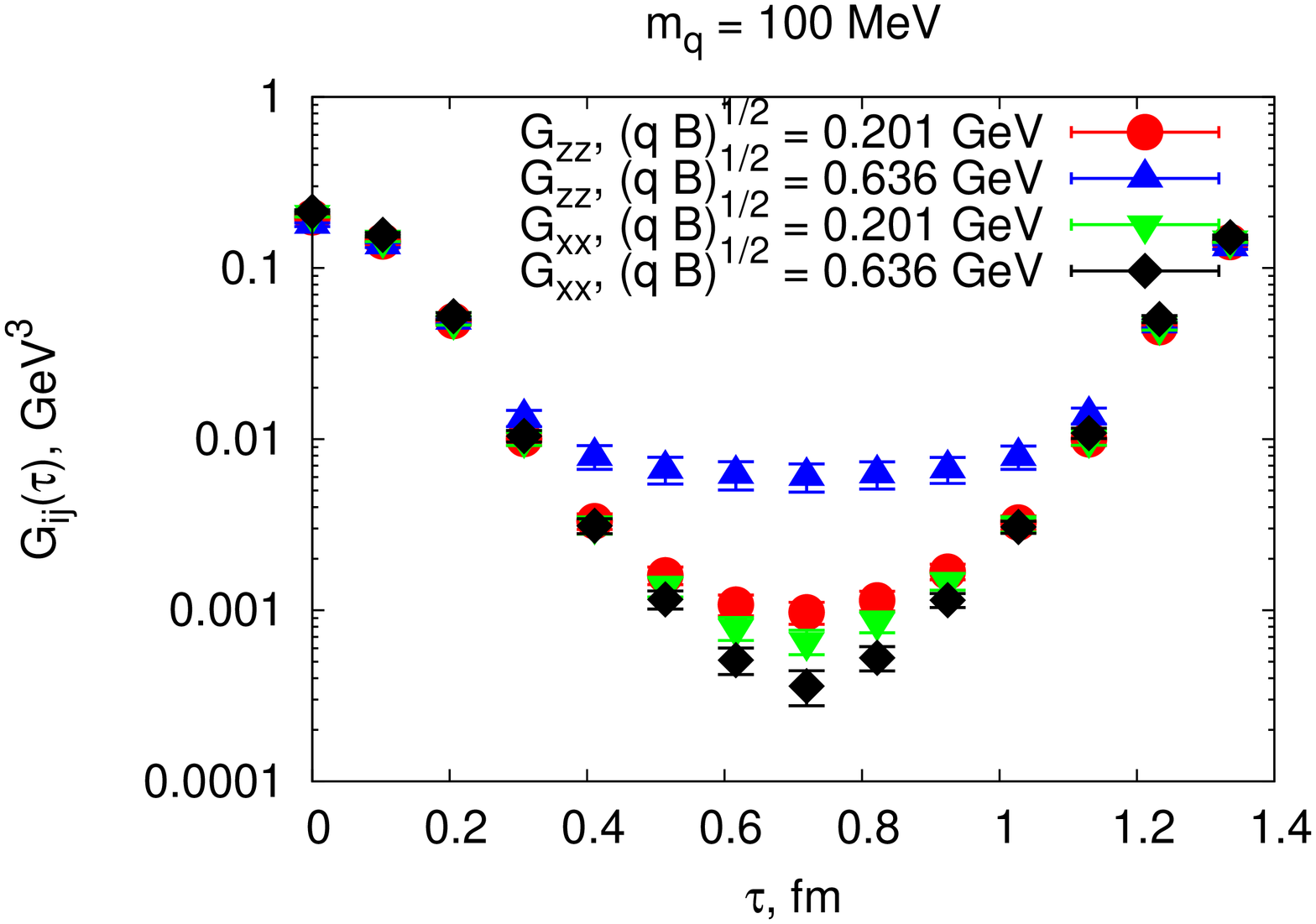}
  \includegraphics[width=5cm, angle=-90]{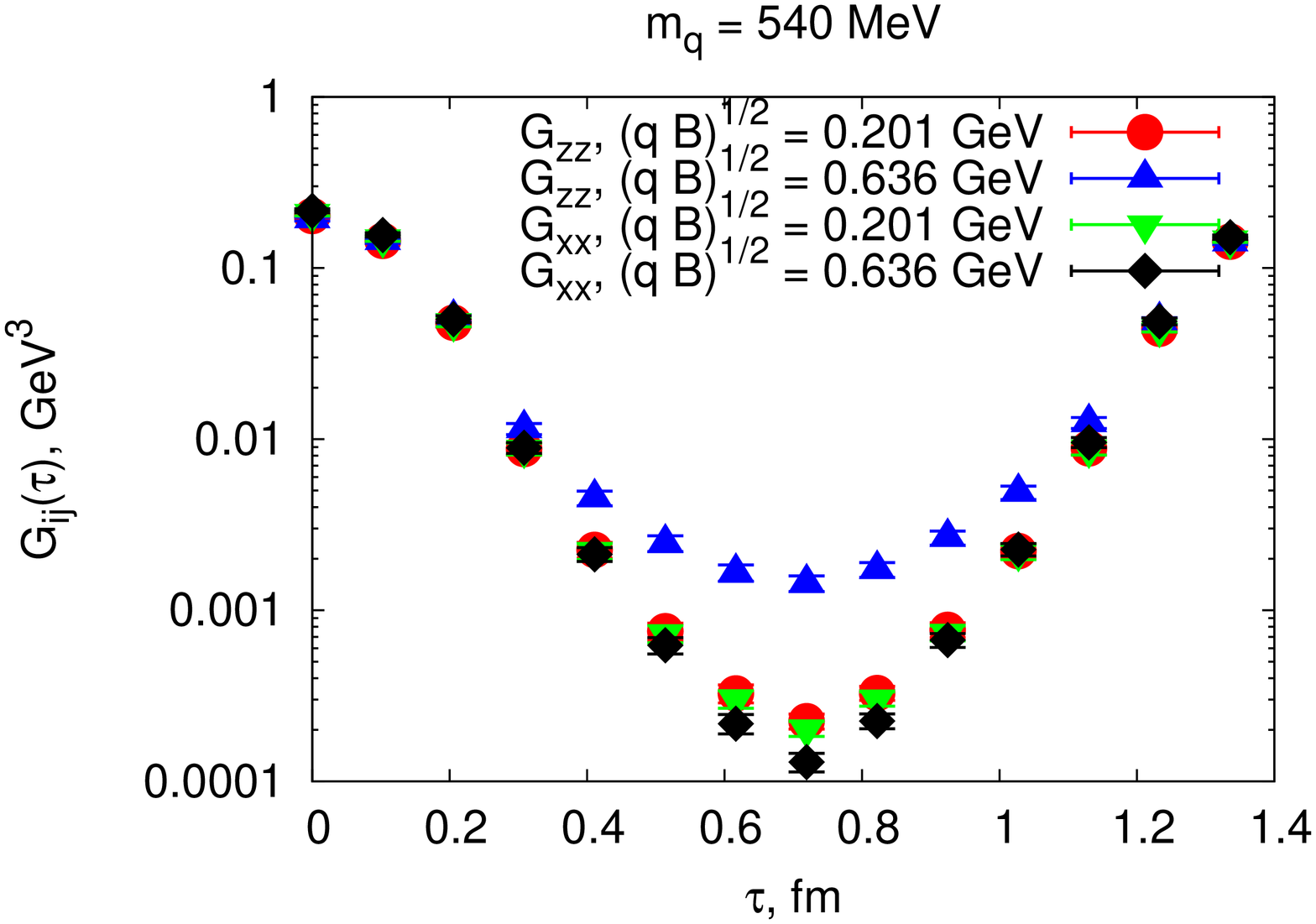}\\
  \caption{Current-current correlators at different quark masses. Above on the left: correlators of the longitudinal components of the current at $\sqrt{q B} = 0.45 \, GeV$ and different quark masses. Above on the right: correlators of the longitudinal and transverse components of the current at $\sqrt{q B} = 0.45 \, GeV$ and different quark masses. Below: correlators of the longitudinal and transverse components of the current at $m_q = 100 \, MeV$ (on the left) and at $m_q = 540\, MeV$ (on the right) and different magnetic fields.}
  \label{fig:correlators}
\end{figure*}

 The correlators $G_{ij}\lr{\tau}$ for various components of the currents and for different magnetic fields and quark masses are plotted on Fig. \ref{fig:correlators}. As in the case of small quark mass \cite{Buividovich:10:1}, at larger quark masses the correlators of the components of the current parallel to the magnetic field decay much slower than for the transverse components. As the bare quark mass $m_{q}$ increases, the correlators also decay faster, however, relative asymmetry between longitudinal and transverse components remains significant.

\begin{figure}
  \includegraphics[width=6cm, angle=-90]{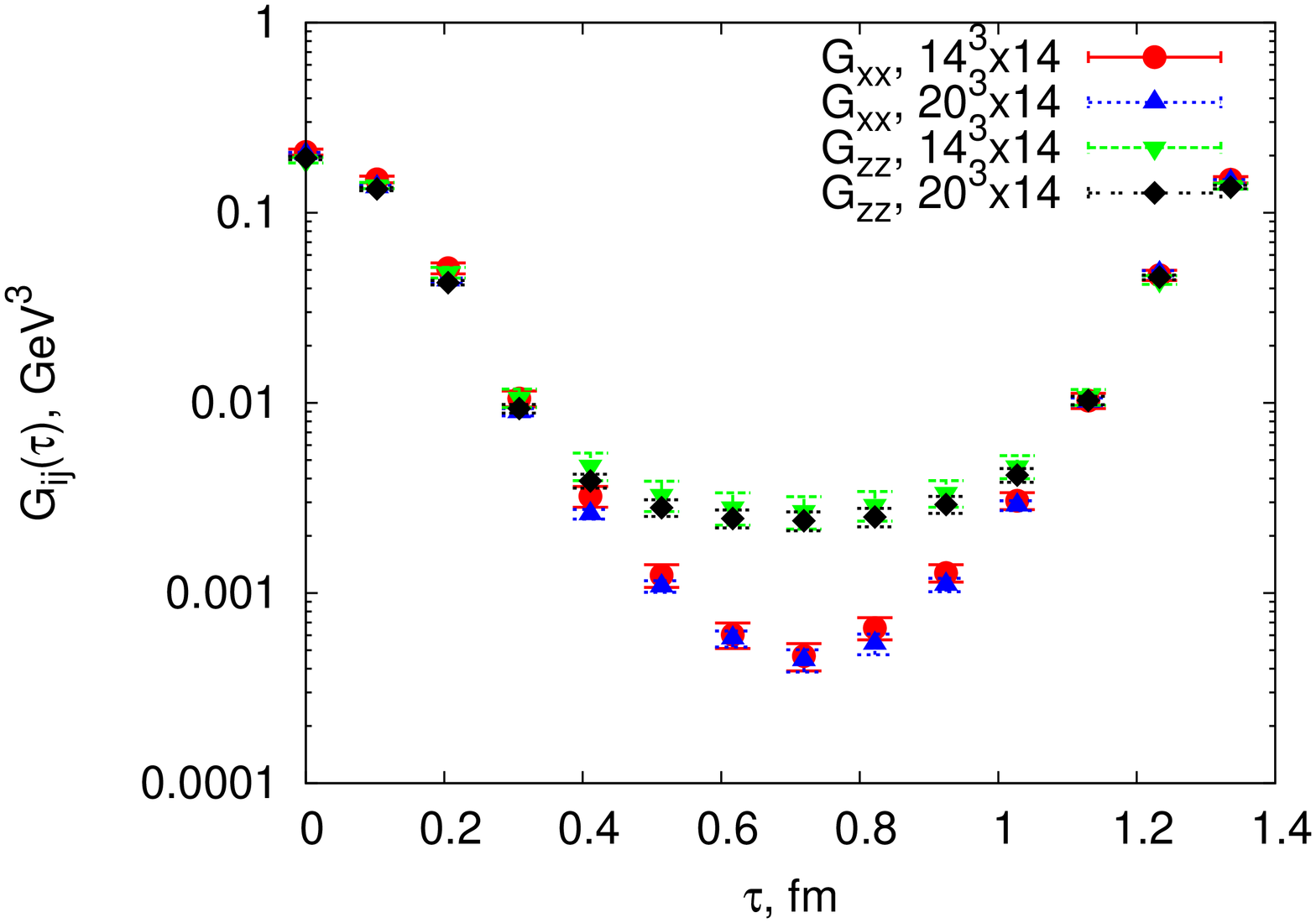}\\
  \caption{A comparison of current-current correlators (\ref{corr_def}) calculated at $14^3\times14$ and $20^3\times14$ lattices at equal values of the lattice spacing and magnetic field strength $\sqrt{q B} = 0.45 \, GeV$.}
  \label{fig:correlator_compare}
\end{figure}

\begin{figure}[h]
  \includegraphics[width=5cm, angle=-90]{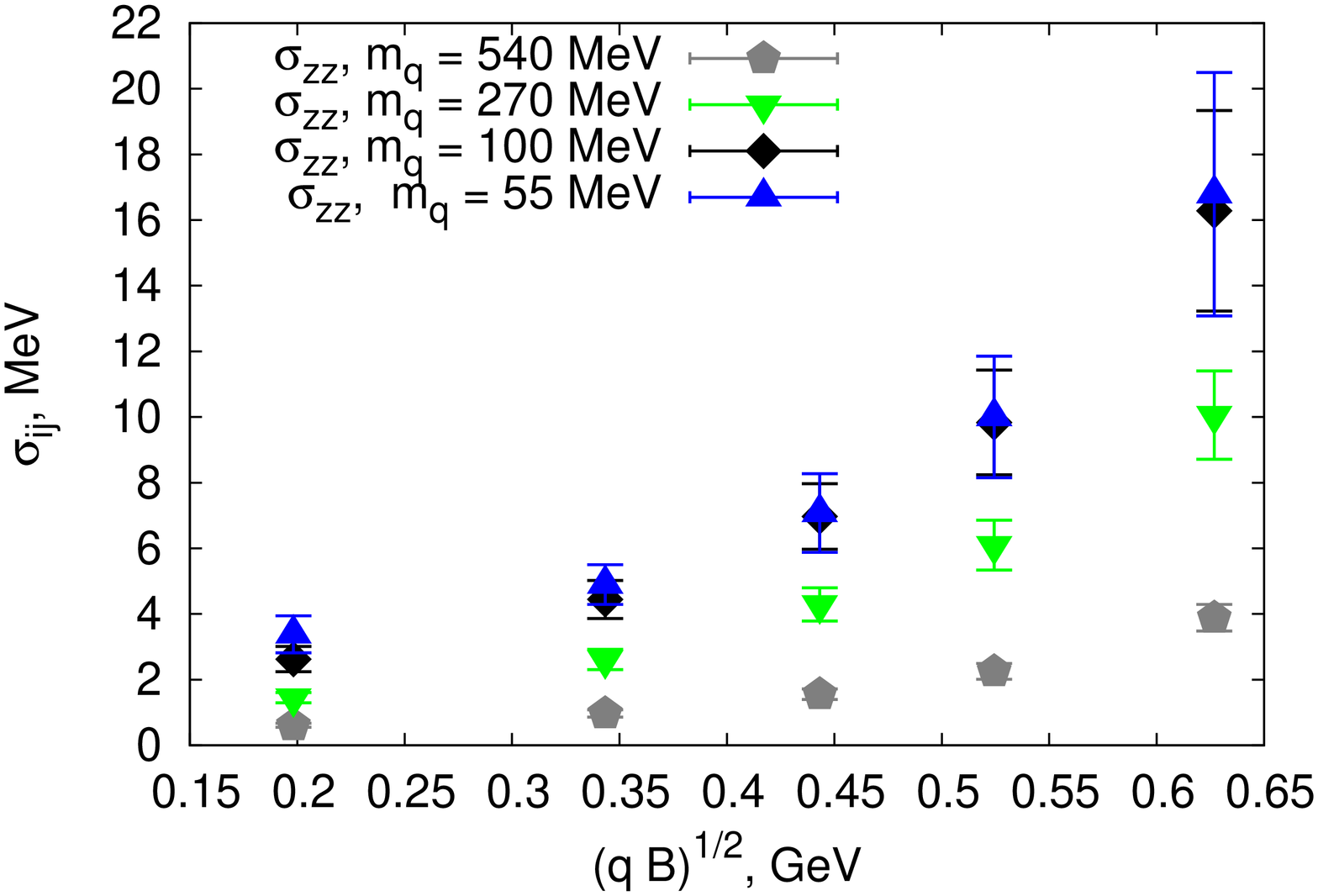}\\
  \caption{Induced conductivity in the direction of the magnetic field as a function of field strength at different quark masses.}
  \label{fig:conductivity_vs_field}
\end{figure}

 A remark is in order concerning the applicability of the expressions (\ref{corr_def}) and (\ref{Kubo_formula}) for the lattices which have equal temporal and spatial extend. While such lattices correspond to zero temperature in the standard lattice lore, one can still apply the expressions (\ref{corr_def}), (\ref{Kubo_formula}) with a small but finite temperature $T = \lr{N_t a}^{-1} = 140 \, MeV = 0.45 \, T_c$, where $a$ is the lattice spacing. This, however, might lead to quite large finite-volume effects. In order to show that finite volume does not affect significantly the vector-vector correlator (\ref{corr_def}), on Fig. \ref{fig:correlator_compare} we compare the correlators (\ref{corr_def}) calculated on the $14^3 \times 14$ and $20^3 \times 14$ lattices at magnetic field strength $\sqrt{q B} = 0.45 \, GeV$. One can see that the correlators agree within error bars for the currents both parallel and perpendicular to the magnetic field. Therefore one can expect that our numerical results are the same as those obtained on the $14 \times L^3$ lattices with $L \gg 14$. Thus for the correlators of vector currents finite-volume effects are smaller than the statistical errors in our simulations, and there is no need to use large anysotropic lattices.

\begin{figure}[h]
  \includegraphics[width=5cm, angle=-90]{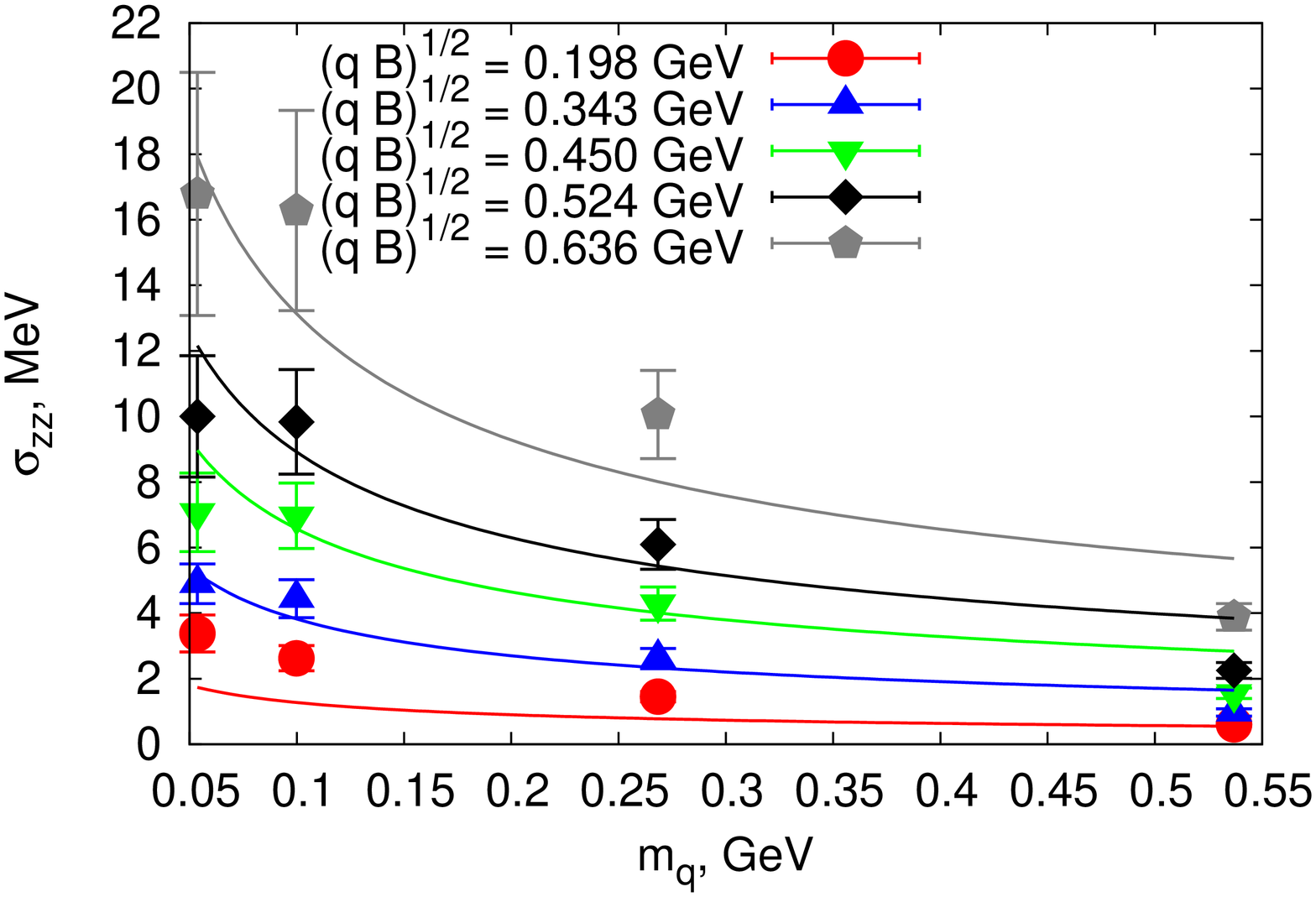}\\
  \caption{Induced conductivity in the direction of the magnetic field as a function of the bare quark mass at different magnetic fields.}
  \label{fig:conductivity_vs_mass}
\end{figure}

 Now we analyze the conductivity extracted form these correlators using the expressions (\ref{corr_def}), (\ref{Kubo_formula}). The dependence of the conductivity in the direction of the magnetic field on field strength and quark mass is illustrated on Figs. \ref{fig:conductivity_vs_field} and \ref{fig:conductivity_vs_mass}. At zero magnetic field the conductivity is equal to zero within error range for all quark masses. As the magnetic field is turned on, the conductivity in the direction of the field grows, but remains equal to zero for transverse directions. For larger quark masses the conductivity in the direction of the magnetic field slowly decreases. Transverse components of the conductivity tensor remain equal to zero.

 In order to describe our data by some phenomenological expression, we have fitted the conductivity in the direction of the field as a function of both field strength and the quark mass by the simplest power-like law:
\begin{eqnarray}
\label{sigma_vs_H_m_fit}
 \sigma_{zz}\lr{m_q, q B} \sim m_{q}^{-\alpha} \lr{|q B|}^{\beta},
\end{eqnarray}
The fit yields the following values of the parameters $\alpha$ and $\beta$:
\begin{eqnarray}
\label{sigma_vs_H_best_fit}
\alpha = \lr{0.45 \, \pm \, 0.06}
\nonumber \\
\beta = \lr{1.1 \, \pm \, 0.2}
\end{eqnarray}

The relative errors of the values of $\alpha$, $\beta$ are within 20\%. We can conclude, therefore, that with a good accuracy the dependence of the induced conductivity on the magnetic field and the quark mass can be described by the following simple expression:

\begin{eqnarray}
\label{sigma_vs_H_m_analytic}
 \sigma_{zz} = \frac{|q B|}{\sqrt{\Lambda m_q}} \, ,
\end{eqnarray}
with some constant $\Lambda$. Such fits are shown on  Fig. \ref{fig:conductivity_vs_mass} by solid lines. It is interesting that the phenomenological expression (\ref{sigma_vs_H_m_analytic}) is non-analytic with respect to the field strength and the quark mass. Moreover, if the conductivity is indeed given by (\ref{sigma_vs_H_m_analytic}), it diverges at zero quark mass.

 Finally, let us relate the conductivity which we have measured to experimental observables. Dilepton emission rate from either cold or hot hadronic matter is given by a general expression \cite{McLerran:85:1, Teryaev:95:1}:
\begin{eqnarray}
\label{DileptonEmissRate}
\frac{R}{V} = -4 e^4 \int \frac{d^3 p_1}{\lr{2 \pi}^3 2 E_1}
\frac{d^3 p_2}{\lr{2 \pi}^3 2 E_2}
L^{\mu\nu}\lr{p_1, p_2}  \frac{\rho_{\mu\nu}\lr{q}}{q^4},
\end{eqnarray}
where $p_1$ and $p_2$ are the momenta of the leptons, $q = p_1 + p_2$,  $L^{\mu\nu} = \lr{\lr{p_1 \cdot p_2 + m^2} \eta^{\mu\nu} - p_1^{\mu} p_2^{\nu} - p_2^{\mu} p_1^{\nu}}$ is the dilepton tensor ($\eta^{\mu\nu}$ is the Minkowski metric), $m$ is the lepton mass. If the electric conductivity is nonzero in the direction of the magnetic field, for sufficiently small $p_1$, $p_2$ one has $\rho_{ij}\lr{q} \approx \sim \sigma_{ij} q/T \sim B_i B_j q/ \lr{|B| T}$. When the mass of the dilepton can be neglected and we are in the rest frame of the dilepton pair, one has $\vec{p_1} = - \vec{p_2} \equiv p \vec{n}$, $q = \lr{2 p, \vec{0}}$ and the spatial components of the dilepton tensor are $L^{ij} = p^{2} \, \lr{ \delta_{ij} -  n_i \, n_j}$. The emission rate is therefore proportional to
\begin{eqnarray}
\label{DileptonEmissRateAnisotr}
\frac{R}{V} \sim \int \frac{d^3 p}{\lr{2 \pi}^3 32 E B p^2 \sqrt{m_q}} \, \lr{\vec{B}^2 - \lr{\vec{B} \cdot \vec{n}}^2}
\sim \nonumber \\ \sim
\frac{B}{\sqrt{m_q}} \, \sin^2\lr{\theta} ,
\end{eqnarray}
where $\theta$ is the angle between the spatial momentum of the outgoing leptons and the magnetic field. Therefore, there should be more soft dileptons emitted perpendicular to the magnetic field than parallel to it. According to (\ref{DileptonEmissRateAnisotr}), the multiplicity of soft leptons should grow linearly with the induced magnetic field. This result, however, should be taken with care, since our data are obtained for quenched theory at sufficiently low temperatures, below the confinement-deconfinement phase transition. In practice, of course, one could also expect numerous soft dileptons coming from the thermalized quark-gluon plasma formed in the collision. Nevertheless, ``anisotropic'' soft dileptons described by (\ref{DileptonEmissRateAnisotr}) could be in principle found in heavy-ion experiments such as RHIC, FAIR or NICA. The mass dependence (\ref{sigma_vs_H_m_analytic}) of the anisotropic part (\ref{DileptonEmissRateAnisotr}) of the lepton production rate could be used, for example, to distinguish between ordinary and strange nuclear matter formed during the heavy-ion collisions. Strangeness enhancement in the vicinity of the confinement-deconfinement phase transition \cite{Rafelski:82:1} might be a relevant situation.

\begin{acknowledgments}
 The authors are grateful to A. S. Gorsky, V.I. Shevchenko, M. Stephanov and A. V. Zayakin for interesting and useful discussions. We are also deeply indebted to D. E. Kharzeev and O. V. Teryaev for very valuable and enlightening remarks on the present work. This work was partly supported by Grants RFBR 09-02-00338-a, RFBR 08-02-00661-a, a grant for the leading scientific schools No. NSh-6260.2010.2 and by the Federal Special-Purpose Programme 'Personnel' of the Russian Ministry of Science and Education. P.B. was partially supported by personal grants from the ``Dynasty'' foundation and from the FAIR-Russia Research Center (FRRC). The calculations were partially done on the MVS 100K at Moscow Joint Supercomputer Center.
\end{acknowledgments}

\bibliography{MyBibliography}
\bibliographystyle{apsrev}

\comment{

}
\end{document}